\documentclass[twoside]{HRIsummer}

\usepackage{graphicx}
\usepackage{amsmath}
\usepackage{amsfonts}
\usepackage{amssymb}
\usepackage{url}
\usepackage[toc,page,title]{appendix}

\title{USING MHD SIMULATIONS TO MODEL H$\alpha$ AND UV SPECTRAL LINES for interpretation of IRIS and NST data}

\author{VIACHESLAV SADYKOV$^{1-4}$, ALEXANDER G. KOSOVICHEV$^{1,2,5}$}

\shortauthors{Sadykov and Kosovichev}

\shorttitle{Modeling of H$\alpha$ and IRIS Lines}

\address{$^{1}$Big Bear Solar Observatory, New Jersey Institute of Technology\\
  40386 North Shore Lane, Big Bear City, CA 92314-9672, USA\\
 e-mail: sasha@bbso.njit.edu, web page: \url{http://www.bbso.njit.edu/}
 \and
$^{2}$NASA Ames Research Center\\ Moffett Field, CA 94035, USA\\
 web page: \url{http://www.nasa.gov/centers/ames/home/}
\and
$^{3}$Space Research Institute (IKI) of Russian Academy of Sciences\\
  Profsoyuznaya Str. 84/32, 117997, Moscow, Russia\\
 e-mail: isharykin@bbso.njit.edu, web page: \url{http://www.iki.rssi.ru/eng/}
 \and
$^{4}$Moscow Institute of Physics and Technology (MIPT)\\
    Institutskiy per. 9, 141700,  Dolgoprudny, Moscow Region, Russia\\
 e-mail: slavasadykov@bbso.njit.edu, web page: \url{http://mipt.ru/en/}
 \and
$^{5}$Stanford University\\
  Stanford, CA 94305, USA\\
 web page: \url{http://sun.stanford.edu/}
 }


\abstract{We present results of non-LTE modeling of H$\alpha$ 6563\,{\AA} and Mg\,II\,k\&h 2796\,{\AA} and 2803\,{\AA} lines. This modeling is important for interpretation of coordinated observations from the recently launched NASA's IRIS mission and from the New Solar Telescope at Big Bear Solar Observatory. Among available codes for the non-LTE modeling, the RH code \cite{Uitenbroek01} is chosen as the most appropriate for modeling of the line profiles. The most suitable Hydrogen and Magnesium atomic models are selected by performing several tests of the code. The influence of the ionization degree on the line profiles is also studied. Radiative-MHD simulations of the solar atmosphere, obtained with the Bifrost code \cite{Gudiksen11}, are used as input data for calculation of synthetic spectra of the H$\alpha$ and Mg\,II\,h\&k lines for particular locations evolving with time. The spectral line variations reveal the presence of MHD waves in the simulation results. We construct oscillation power spectra of the line intensity for different wavelength, and compare these with the corresponding height-dependent power spectra of atmospheric parameters from the simulations. We find correlations between the power spectra of intensities of the line profiles at certain wavelengths and the power spectra of the atmospheric parameters at the tau-unity heights for these wavelengths. These results provide a new diagnostic method of chromospheric oscillations; however, larger amounts of data are needed to confirm these correlations.}

\begin{document}
%\maketitle

\section{INTRODUCTION}

Profiles of solar spectral lines and their behaviour have been studied for many years. However, many questions about factors responsible for their formation and line shape are still being discussed. These questions are among of the most important and interesting in solar physics because the spectral lines contain information about the structure of the solar atmosphere (its physical parameters, e.g. the composition, temperature distribution, magnetic fields, velocities etc.), and also the information about dynamic processes in the atmosphere (such as oscillations, shocks, and eruptions). Different lines are formed at different heights of the atmosphere. Therefore, simultaneous observations of different spectral lines can provide us with the information about different layers of the solar atmosphere.

\begin{figure}[t]
\centering
\includegraphics[width=1.0\linewidth]{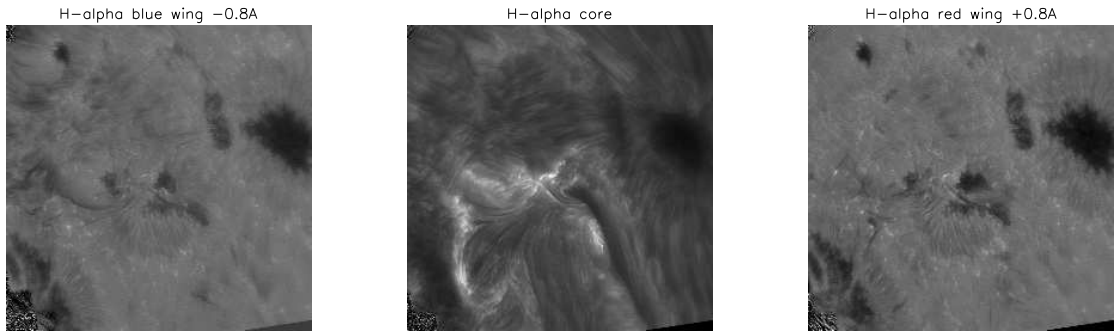}
\caption{The NST/VIS observations of a flaring region obtained on 12 June, 2014, 20:58 UT, with three different filters: H$\alpha$ blue wing -0.8{\AA} (left panel), H$\alpha$ line core (center panel), H$\alpha$ red wing +0.8\AA (right panel)}
\label{figure1}
\end{figure}

However, the spectroscopic data are very complicated even for individual lines because of the multi-scale dynamical structure of the atmosphere. Figure~\ref{figure1} displays images obtained with the Visible Imaging Spectrometer (VIS) \cite{Cao10} at the New Solar Telescope (NST), Big Bear Solar Observatory, in three different wavelengths across the H$\alpha$ line. The 1.6\,m NST is currently the world-largest solar telescope. Figure~\ref{figure1} clearly shows that different parts of the H$\alpha$ line effectively sample different heights of the atmosphere. In the line wings the upper photospheric layers are observed, and in the core of the H$\alpha$ line one can see fibrils~--- structures typical for the chromosphere. Therefore, even single spectral line contains information about different heights of the solar atmosphere.

The most interesting question of the solar imaging spectroscopy is how to derive physical parameters of the atmosphere from the line profile measurements. In general, this is an inverse ill-posed problem. To solve the inverse problem it is necessary to study the forward problem, and find correlations between variations of the line profile shape and physical parameters of the atmosphere. For this we need to solve the radiation transfer equation for realistic solar conditions.

It is possible to observe line spectra with very high spatial, temporal and spectral resolutions. However, results of the line profile modeling still remain unique products even for the forward problem. Thus, it is necessary to understand what kind of information we can obtain from spectra itself, without solving inverse problem for each line profile. Modeling can help us to determine some characteristic behavior of spectra and corresponded behavior of the atmosphere.

In this paper we describe the initial experience and results obtained with the currently available numerical radiation transfer codes. Our purpose is calculation of the H$\alpha$ and Mg\,II\,h\&k line profiles. In Section~2 we present analysis of the radiation transfer codes. In Section~3 we describe computational set up, input parameters and tests of the RH code \cite{Uitenbroek01} which we found most suitable for our purpose. In Sections~4 and~5 we describe calculations of the line profiles, and correlations between the atmospheric wave dynamics and variations of the line profiles. In Appendix we present a mathematical description of the RH approach, collected from various sources.

\section{ANALYSIS OF RADIATION TRANSFER CODES}

Our analysis is focused on the H$\alpha$ and Mg\,II\,h\&k lines which are among most important spectral lines for diagnostics of the solar chromosphere. The H$\alpha$ line is observed with many instruments including high-resolution observations with the VIS instrument at NST \cite{Cao10}. The Mg\,II\,h\&k lines are currently observed with high spatial and temporal resolution with the NASA's IRIS mission \cite{DePontieu14}. These lines are usually optically thick, and scattering plays a key role in their formation process. It is impossible to correctly model these lines under the local thermodynamic equilibrium (LTE) consideration. Only the radiative transfer codes which include non-LTE effects can be applicable for modeling these lines.

We considered three non-LTE codes: PANDORA \cite{Avrett92}, MULTI\,2.3 and RH \cite{Uitenbroek01}. The PANDORA radiative transfer code \cite{Avrett92} is available from the web site \url{https://www.cfa.harvard.edu/~avrett/pandora.html}. It solves the non-LTE radiation transfer problem in a one-dimensional atmosphere and allows to include many different effects like the partial frequency redistribution effect (PRD), and also effects caused by deviations from the statistical equilibrium. We successfully compiled the code and ran the demos and tests. Because of the complexity, our work on setting up this code for the IRIS and NST data is not yet completed. Applications of this code to modeling the IRIS's data was recently presented by Prof. Avrett~\cite{Avrett13}.

The MULTI\,2.3 radiation transfer code is available from the web page \url{http://folk.uio.no/matsc/mul23/}. This code has well-organised documentation, and it was easy to configure it for the H$\alpha$ line modeling. However, MULTI\,2.3 does not include the partial redistribution (PRD) effects for the line formation process. Also, this code does not allow us to calculate the lines, the shape of which cannot be approximated by a single Gaussian or a Voigt profile, e.g Mg\,II\,h\&k lines (Fig.~\ref{figure2}) in moving atmospheres. These simplifications are not important for modeling of the H$\alpha$ line because the PRD effects are not strong (Figure 5 of Leenaarts et al \cite{Leenaarts12}). However, the PRD effects play a key role in the Mg\,II\,h\&k line (Figure 11 of Leenaarts et al \cite{Leenaarts13a}). Therefore, this code is not suitable for modeling the IRIS data.

The MPI-parallelized version of the RH radiative transfer code is available on the web site \url{http://iris.lmsal.com/software.html}, and suggested for modeling of the IRIS data. We thank Han Uitanbroek for providing us with the non-parallelized version of the RH code. The RH algorithm originally developed by G. Rybicki and D. Hummer \cite{Rybicki91}, \cite{Rybicki92} uses the MALI iterative scheme (Multilevel Accelerated Lambda Iteration), which was updated by Uitenbroek and includes the PRD effects \cite{Uitenbroek01}. This update allows the users to calculate the Mg\,II\,h\&k line profiles which are a very powerful instrument for diagnostics of the upper chromosphere \cite{DePontieu14}. The RH code uses some simplifications; and one of the assumptions is that the ionization degree if fixed during the computational process. The ionization degree can be provided as an input parameter, or it can be calculated in the LTE approximation. The statistical equilibrium is assumed for the population of atomic levels. This code provides the opportunity to calculate the H$\alpha$ and the Mg\,II\,h\&k line profiles simultaneously, and thus we decided to use it in this work.

\section{PARAMETER SETTINGS FOR THE RH RADIATIVE TRANSFER CODE}

In this section we describe parameters and settings which we use for the RH code. The input data: the temperature, total electron number density, column mass, and velocity profiles along the line of sight are adopted from the 3D MHD code Bifrost.

The RH code uses atomic models of a specific format. The variety of different atomic models in the default RH version is sufficient to calculate many photospheric and chromospheric lines. In addition to the H$\alpha$ and Mg\,II\,h\&k lines, we successfully calculated C\,II\,1334/1335{\AA} lines, Fe 6173\,{\AA} and Fe\,6302/6303\,{\AA} lines. Different atomic models for the same element are included in the RH distribution. The models usually differ from each other by a number of included atomic levels and transitions. More detailed atomic models provide more precise line profiles, but use more computing time. Therefore, one of our tasks was to find the simplest atomic model which gives sufficiently accurate results.

\begin{figure}[t]
\centering
\includegraphics[width=0.95\linewidth]{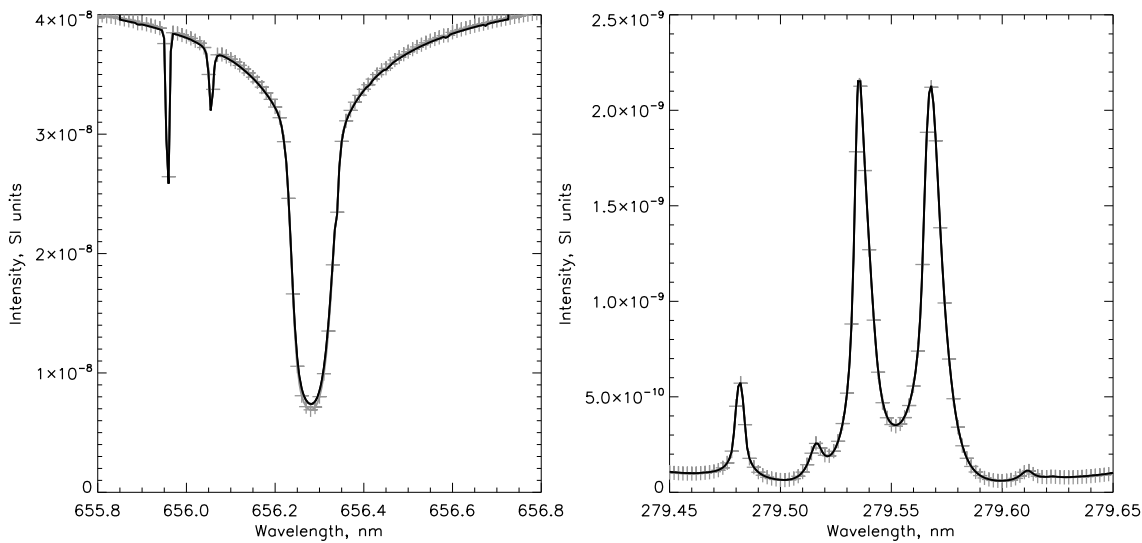}
\caption{Comparison of the line profiles obtained with different atomic models in the RH. Left panel: H$\alpha$ line profile obtained with ``H\_6'' atomic model (black solid curve) and with ``H\_9'' atomic model (gray crosses). Right panel: Mg\,II\,k line profiles obtained with ``MgI+II'' atomic model (black solid curve) and ``MgII'' atomic model (gray crosses).}
\label{figure2}
\end{figure}

The left panel in Figure~\ref{figure2} shows the H$\alpha$ line profiles obtained with two different models of Hydrogen. The first model includes the first five atomic levels and one ionization state, and is labeled as ``H\_6''. The second model (``H\_9'') includes 9 levels and one ionization state. The FALC model \cite{Fontenla93} of the solar atmosphere from the RH distribution was used in these calculations. The PRD effects are included in these tests although these effects are not very important for the H$\alpha$ line. Figure~\ref{figure2} displays that the greatest difference between the ``H\_6'' and ``H\_9'' models is in the H$\alpha$ core and negligible in the line wings. For further calculations we use the ``H\_9'' model.

The right panel in Figure~\ref{figure2} displays the Mg\,II\,k line profiles obtained with two different Magnesium models: 1) ``MgI+II'' atomic model, which includes 56 levels of Mg\,I and 10 levels of Mg\,II, and also ground level of Mg\,III, 2) ``MgII'' atomic model, which includes only 10 levels of Mg\,II and the ground level of Mg\,III. Since the difference between the line profiles of these models is insignificant, but computations for ``MgII'' are three times faster, we decided to use this model. Therefore, we decided to use the pair of atomic models ``H\_9'' and ``MgII'' for modeling the H$\alpha$ and the Mg\,II\,h\&k line profiles.

\begin{figure}[t]
\centering
\includegraphics[width=0.95\linewidth]{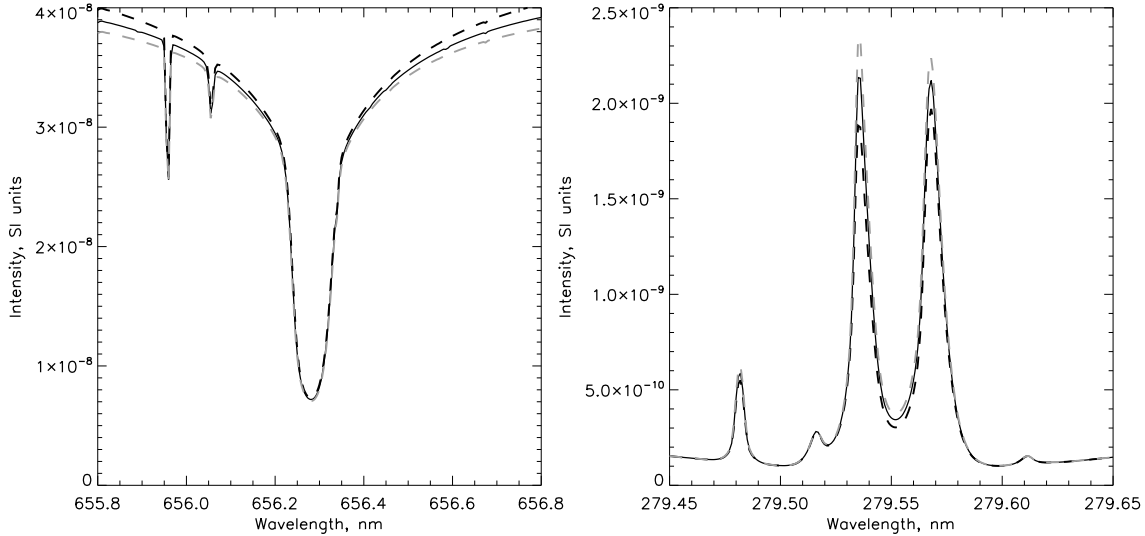}
\caption{Comparison of line profile changes due to changes of the ionization degree. The black solid line profiles are obtained for the FALC atmospheric model. The grey dashed line profiles are obtained for the ionization degree 25\% higher, and the black dashed profiles are for the ionization degree 25\% lower than in the FALC model. The left panel corresponds to the H$\alpha$ line, the right panel corresponds to the Mg\,II\,k line.}
\label{figure3}
\end{figure}

We mentioned before that one of the simplifications of the RH code is that the ionization degree of elements is fixed during the computational process. The ionization degree can be provided as an input parameter or calculated in the LTE approximation. Thus, it is important to investigate how changes of the ionization affect the line profiles. Figure~\ref{figure3} illustrates sensitivity of the line profile to changes of the ionization degrees by $\pm 25\%$ relative to the FALC model. One can see that the significant changes of the H$\alpha$ line profile take place only in the H$\alpha$ wings. However, the Mg\,II\,k line is more sensitive to changes of the ionization degree. Its intensity profile changed significantly, by more than 10\% for these changes of ionization. The strongest changes of the Mg\,II\,k line profile take place in the line peaks and in the dip between peaks. Thus, accurate calculations of the ionization degree are particularly important for modeling the Mg\,II lines.

\section{STUDY OF MHD WAVES AND THEIR APPEARANCE IN LINE PROFILES}

To investigate how dynamical processes in the solar atmosphere are reflected in the line profile variations we applied the RH code to the simulation results obtained with the Bifrost code. The BIFROST is a state-of-the-art Radiative-MHD simulation code \cite{Gudiksen11} aimed at the realistic modeling of solar and stellar atmospheres. Some results of the BIFROST simulation are publicly available from the web site: \url{http://sdc.uio.no/search/simulations}. The results are obtained for the simulation domain of 24\,x\,24\,x\,17\,Mm$^3$ with 48\,km horizontal resolution and 19-100\,km vertical resolution. The vertical grid is non-uniform. The available simulation results represent 156 snapshots of atmospheric layers from the upper convective zone to the corona with 10 second cadence. Magnetic field is represented by a weak bipolar magnetic structure with average field strength of $\sim$50\,G. Information about the density, velocity, magnetic field, internal energy, electron number density, gas pressure and temperature is available for each snapshot of this simulation. The results of the BIFROST simulations are already used in several papers to study the behavior of the H$\alpha$ and Mg\,II line profiles (Leenaarts et al \cite{Leenaarts12}, \cite{Leenaarts13a}, \cite{Leenaarts13b}).

The spatial resolution of the BIFROST code (48\,km) corresponds to 0.067 arcseconds for observation at the disk center. This is 5 times smaller than the spatial resolution of the IRIS, which is $\sim$0.33 arcseconds. To compare our results with IRIS observations and also to smooth sharp gradients in the atmosphere to make the code more stable, we averaged the simulation data over 5\,x\,5 grid points in the horizontal plane. We used averaged temperature, vertical velocity, column mass density and electron number density as the input data for the RH code, and calculated the evolution of the H$\alpha$ and the Mg\,II\,h\&k line profiles for selected columns.

\begin{figure}[t]
\centering
\includegraphics[width=0.95\linewidth]{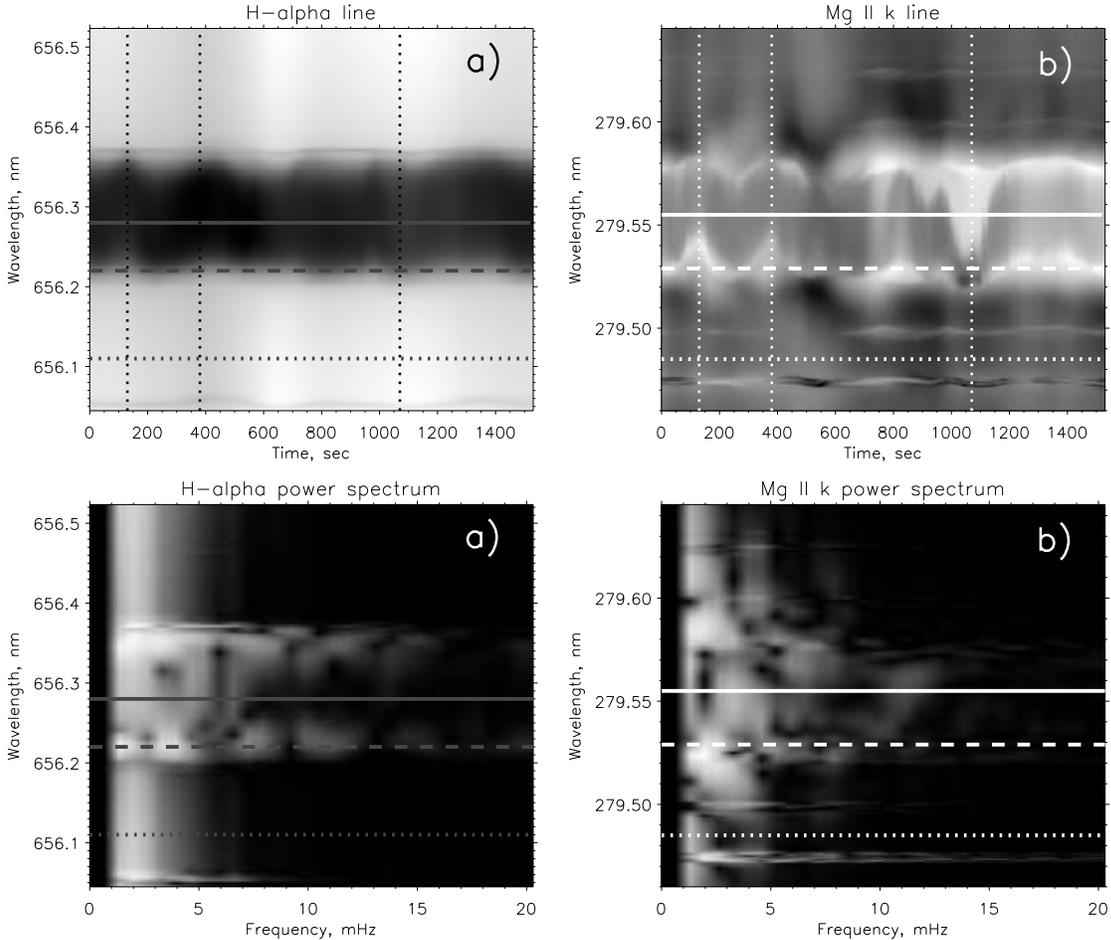}
\caption{The spectrograms of the H$\alpha$ and Mg\,II\,k lines and their power spectra: a) H$\alpha$ line; b) Mg\,II\,k line. Horizontal solid lines show the line center, dashed lines show the selected near wing points, and dotted lines indicate the far wings (continuum). The power spectra are displayed in logarithmic scale.}
\label{figure4}
\end{figure}

\begin{figure}[t!]
\centering
\includegraphics[width=0.72\linewidth]{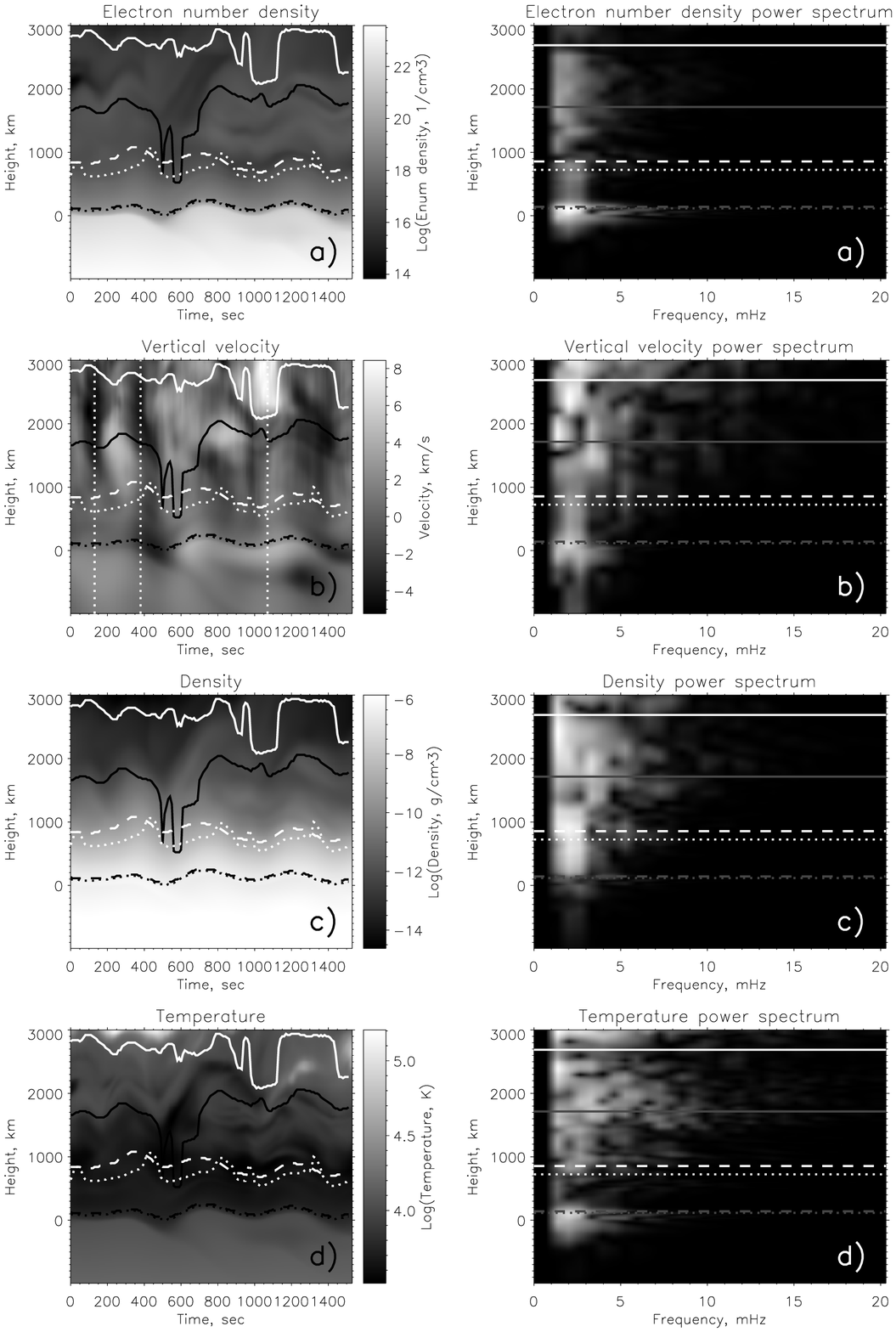}
\caption{Time-height diagrams and power spectra of atmospheric parameters: a) electron number density; b) vertical velocity; c) density; and d) temperature. The curves plotted over the atmospheric parameters correspond to the optical $\tau = 1$ levels for the wavelengths indicated in Figure 4 and plotted with the same style of line. The horizontal lines plotted over the power spectra images determine the average optical $\tau = 1$ level (the effective formation zone) for the wavelengths which we selected on Figure 4 and plotted with the same style. Black(grey) lines correspond to the H$\alpha$ line, white lines correspond to the Mg\,II\,k line. The vertical white dotted lines in the velocity diagram indicate the upflow/downflow events discussed in the text. The power spectra are displayed in logarithmic scale.}
\label{figure5}
\end{figure}

The top panels in Figure~\ref{figure4} display spectrograms as a function of time and wavelength for the H$\alpha$ and the Mg\,II\,k line profiles. The left panels of Figure~\ref{figure5} show the evolution of the input parameters in a selected atmospheric column with time. The presence of waves is evident in both figures, thus, the waves propagating in the atmosphere manifest themselves in the line profile variations. The waves are excited by the turbulent convection in shallow subphotospheric layers \cite{Kitiashvili11}, \cite{Stein01} and travel into the atmosphere where they develop shocks.

The estimated periods of these waves are around 3-8 minutes, which is of the same order as the period of oscillations of mottles and loops (5-10 minutes) in the quite Sun network region \cite{RouppeVanDerVoort07}, and the acoustic oscillations of the photosphere (the dominant period is about 5 minutes). The fact that the oscillations appear not only in the atmospheric properties, but also in the spectrograms, provides us with an opportunity to investigate oscillatory characteristics of different atmospheric layers from observations of the line profiles.

The bottom panels in Figure~\ref{figure4} and the right panels in Figure~\ref{figure5} display the oscillation power spectra. Figure~\ref{figure4} shows the power spectra of the intensity of the H$\alpha$ line (panel a) and the Mg\,II\,k line (panel b) as a function of oscillation frequency and wavelength. Figure~\ref{figure5} shows the power spectra for the electron number density (panel a)), vertical velocity (panel b)), density (panel c)) and temperature (panel d)) as a function of frequency and height in the solar atmosphere. We used the Fast Fourier Transform with the Hanning window to obtain all these spectra. The spectral resolution of each power spectrum is around 0.7\,mHz. We cut the first two harmonics for better contrast of the displayed spectra. We selected three wavelength points corresponding to the line core wing and continuum for each H$\alpha$ and Mg\,II\,k lines for a detailed study. In Figure~\ref{figure5} we indicate the $\tau = 1$ levels corresponding to the selected line-profile points in the images of Fig.~\ref{figure4} showing variations of the atmospheric properties.

\begin{figure}[t]
\centering
\includegraphics[width=0.75\linewidth]{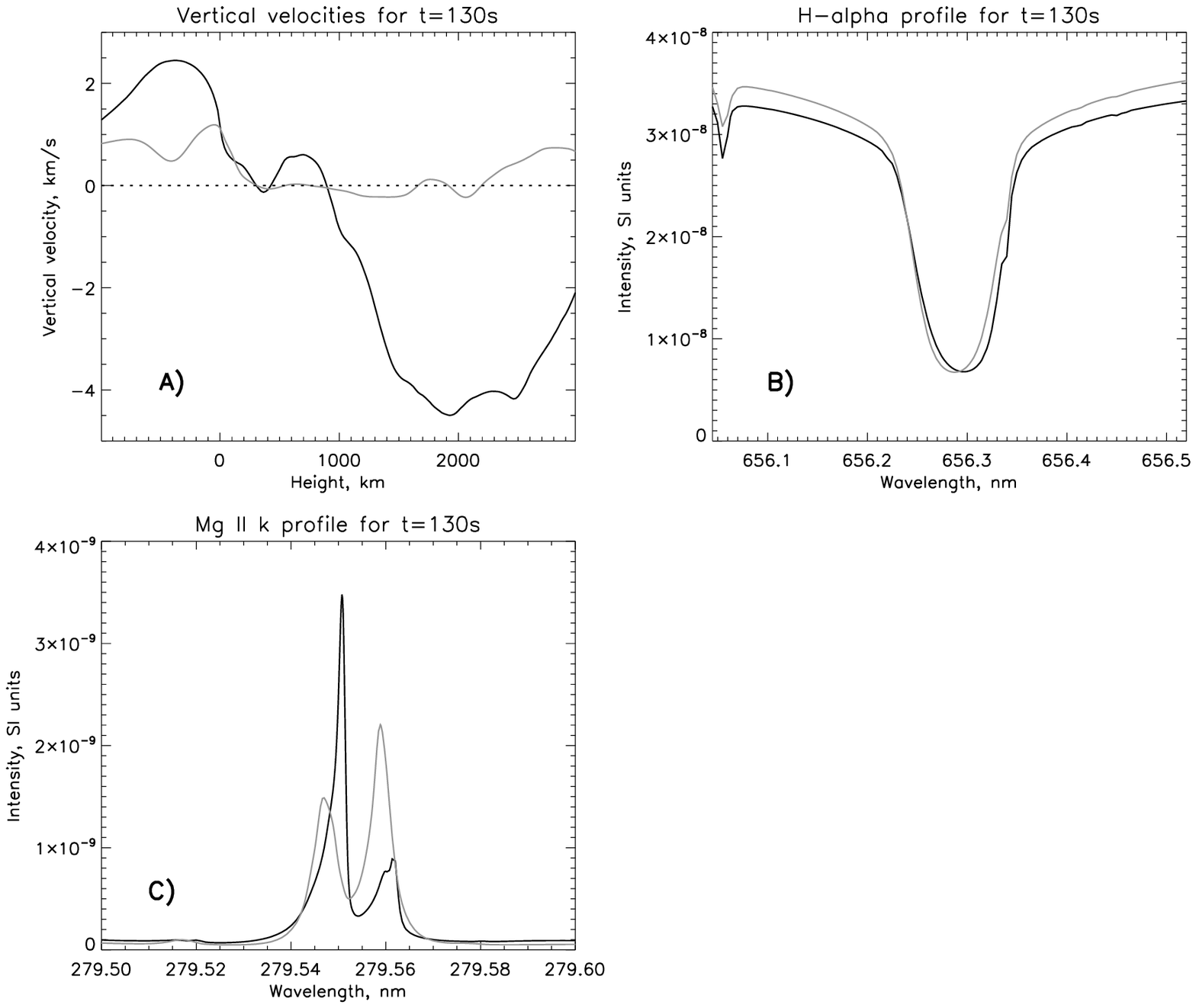}
\caption{Characteristics of the chromospheric downflow event at t=180\,s: A) vertical velocity; B) H$\alpha$ line profile; C) Mg\,II\,k line profile. The grey curves show the mean line profiles averaged over the data set}
\label{figure6}
\end{figure}

\begin{figure}[t]
\centering
\includegraphics[width=0.75\linewidth]{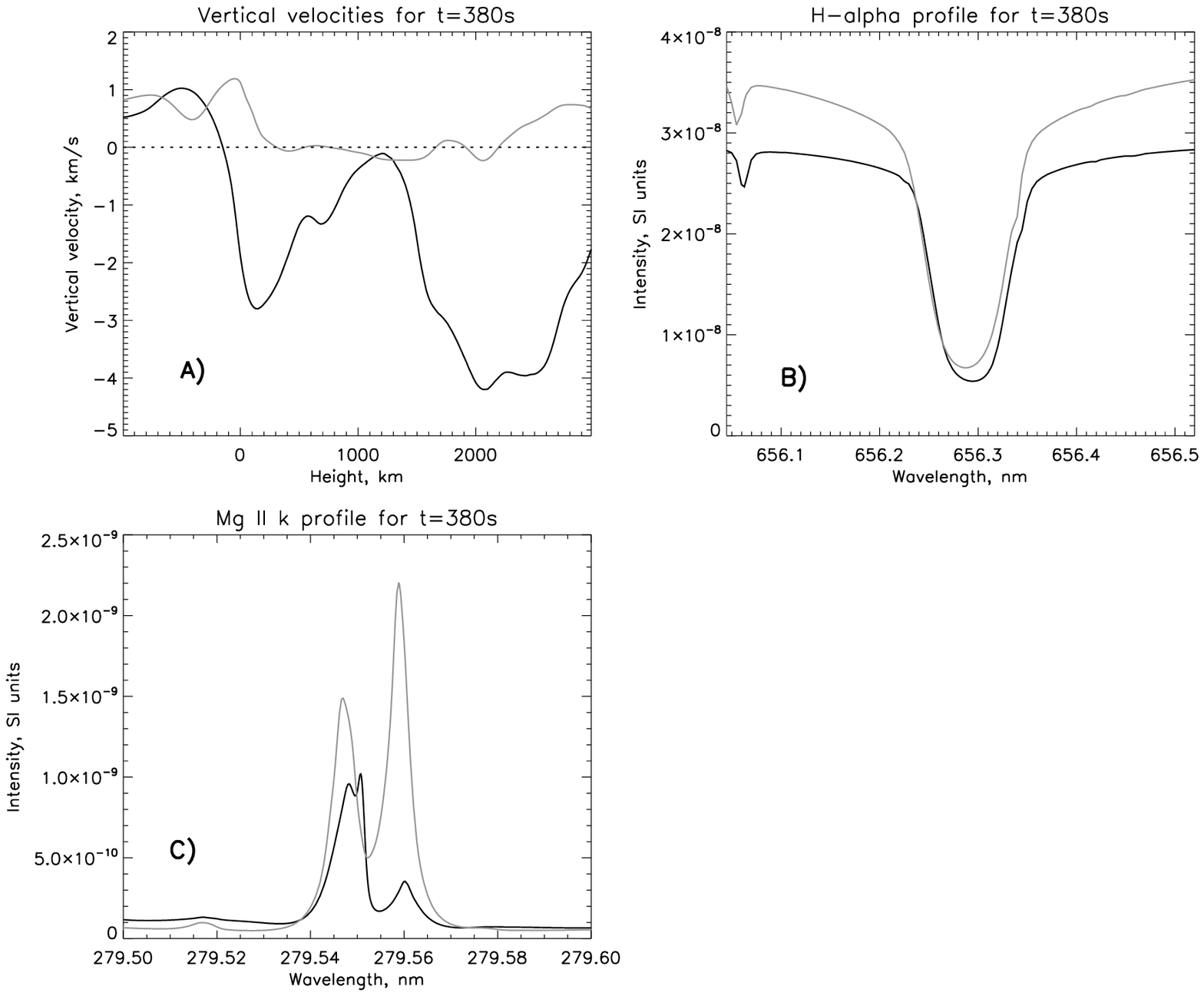}
\caption{Characteristics of the chromospheric downflow event at t=380\,s: A) vertical velocity; B) H$\alpha$ line profile; C) Mg\,II\,k line profile. The grey curves show the mean line profiles averaged over the data set}
\label{figure7}
\end{figure}

\begin{figure}[t]
\centering
\includegraphics[width=0.75\linewidth]{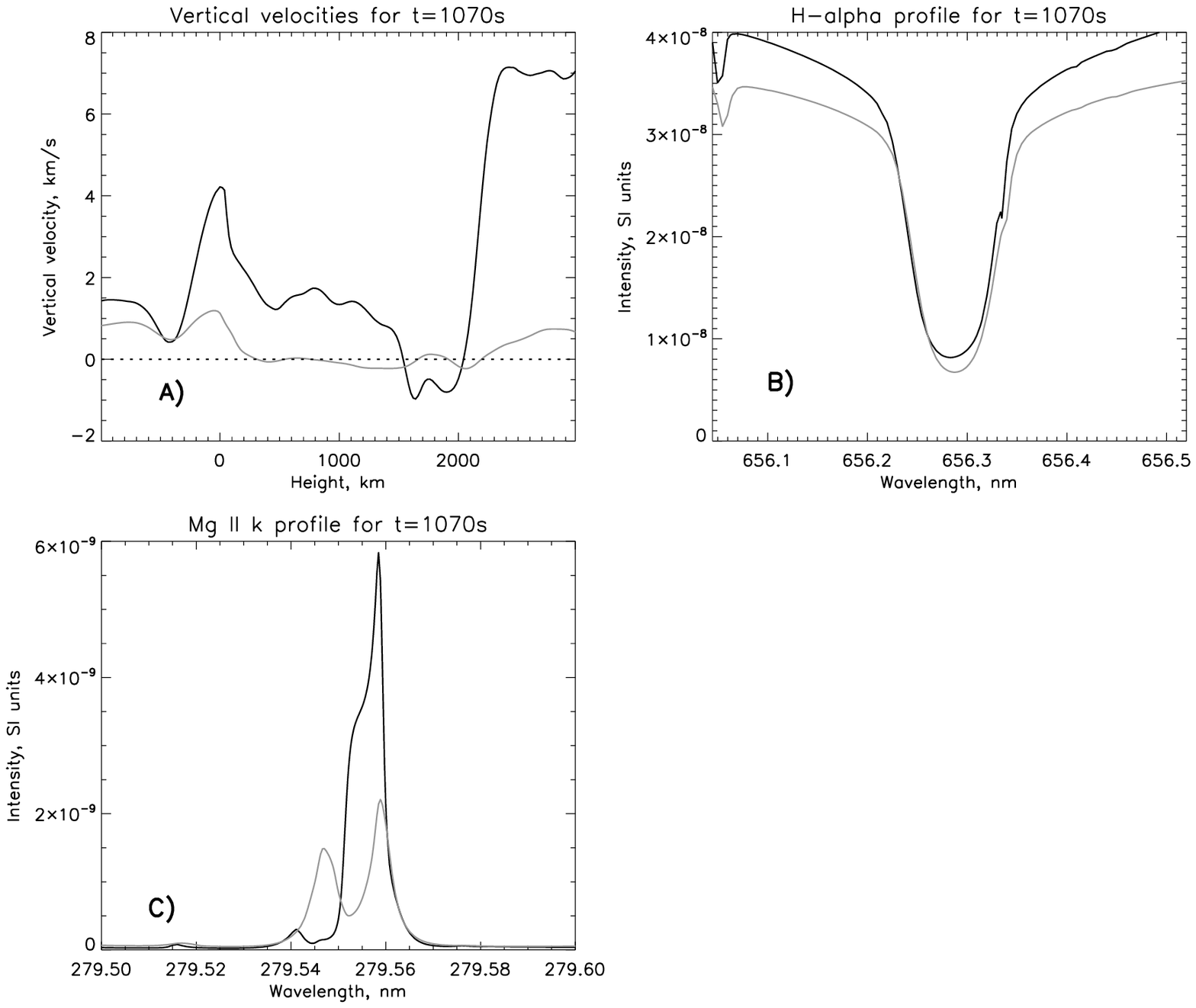}
\caption{Characteristics of the chromospheric upflow event at t=1070\,s: A) vertical velocity; B) H$\alpha$ line profile; C) Mg\,II\,k line profile. The grey curves show the mean line profiles averaged over the data set}
\label{figure8}
\end{figure}

The brightest atmospheric events which are also reflected in the line spectra are strong upflows and downflows seen in the modeled velocity variation. We selected two upflows and one downflow in the upper chromospheric region and decided to consider line profiles for these events in details. The downflows correspond to $t=130$\,sec and $380$\,sec, and the upflow event corresponds $t=1070$\,sec. These events are marked with vertical white dashed lines in the velocity variation in Fig.~\ref{figure5}b. Figures~\ref{figure6},~\ref{figure7}~and~\ref{figure8} display the velocity profiles, and the H$\alpha$ and Mg\,II\,k line profiles for these moments. For better determination of specific features emerged in the line profiles, we also show the meanline profiles averaged over the whole set.

In the next section we discuss correlations between the spectra of the atmospheric parameters and the spectra of the synthetic line profiles, and specifically consider manifestation of strong upflows and downflows in the line profiles.

\section{COMPARISON OF LINE SPECTROGRAMS WITH ATMOSPHERIC PARAMETERS}

Figures~\ref{figure4}~and~\ref{figure5} allow us to discuss characteristic correlations between the power spectra of the line spectrograms and the atmospheric parameters.

1. Figure~\ref{figure4} displays the power spectra of the H$\alpha$ line profile and the Mg\,II\,k line profile. Oscillations with the frequency of 2\,mHz are dominant in the H$\alpha$ continuum signal. The 2\,mHz oscillations are also prominent in the power spectra of atmospheric parameters (density, temperature, vertical velocity and electron number density) at the heights corresponding to the optical depth $\tau = 1$ of the H$\alpha$ continuum. For the Mg\,II\,k line continuum the $\tau = 1$ layer is located at heights of about 800\,km. The power spectra of the atmospheric properties become more complicated in the middle chromosphere at heights of about $\sim$1000\,km.

2. The near wing point of the H$\alpha$ line is formed at approximately the same $\tau = 1$ heights as the continuum. However, the power spectrum at this wavelength is more complicated and additional power at $\sim$3.6\,mHz appears. The reason for this behavior may be the following. The main difference of the near wing and continuum is that the near wing points are sensitive not only to intensity oscillations but also to the line-profile Doppler shifts. The superposition of these two effects may lead to presence of additional harmonics in the power spectra.

3. The H$\alpha$ and Mg\,II\,k line core points have more complicated oscillation power spectra than the near wing and continuum points. The high-frequency power at 5-10\,mHz is substantially increased for the line center. The $\tau = 1$ layer for the line core is $\sim$1500\,km higher than for the near wing and continuum points. It is $\sim$1700\,km high for the H$\alpha$ line core and $\sim$2700\,km high for the Mg\,II\,k line center. The power spectra of the atmospheric properties are also very complicated at these heights, and show the increase of the high-frequency power of the vertical velocity, temperature, density, but not of electron density. Perhaps, the high-frequency oscillations on the upper chromosphere are due to non-linear wave effects and shock formation.

4. Figures~\ref{figure6}~and~\ref{figure7} display vertical velocities and line profiles for two downflow events in the upper chromospheric region (above 1000\,km). Both downflows have maximum velocity around -4\,km/s, and appear in the upper chromosphere in the region of formation of the Mg\,II and H$\alpha$ line cores (see $\tau = 1$ layers in Fig.~\ref{figure5}). Both events demonstrate a significant difference between the two peaks of the Mg\,II\,k line: the k2r peak (corresponding to longer wavelengths \cite{Leenaarts13b}) is significantly lower than the k2v peak (corresponding to shorter wavelengths). Two line peaks, and also the line dip are redshifted with respect to the profile averaged over the data set. However, the center-of-gravity of the Mg\,II\,k line is definitely blueshifted for these cases (because the k2v peak is significantly stronger). This fact needs to be taken into account in automatic procedures of line analysis. In such cases, the center-of-gravity is not the best measure of Doppler shifts for the Mg\,II lines. Behavior of the H$\alpha$ line for these downflows is simpler. In both cases, redshift is observed with respect to the averaged profile. Also the H$\alpha$ profile does not change as significantly as the Mg\,II profiles. Figure~\ref{figure8} displays upflow in the upper chromospheric region. This upflow is not so deep as the downflows considered before: it starts at 2000\,km, which is above the $\tau = 1$ layer for the H$\alpha$ line. Figure~\ref{figure4} shows that this upflow is very prominent in the spectra. Contrary to the downflows, the line peaks and dip of the upflow events have significant blueshift with respect to the averaged profile, and the k2r peak is significantly brighter than the k2v peak. Also, as expected, the H$\alpha$ line profile is not strongly affected by the upflow. Correlations for these kinds of the Mg\,II profiles (with two line peaks) have been explored in the paper of Leenaarts et al \cite{Leenaarts13b} who also used the Bifrost simulation results.

The question is how to identify the chromospheric upflows and downflows in the spectrograms. It is hard to identify these from the H$\alpha$ spectra. The H$\alpha$ line is quite stable, and usually not so sensitive to such features. Also it is hard to distinguish these effects from oscillations. However, these features are very prominent in the Mg\,II\,k spectra. Their spectral characteristics are different from oscillations, and indicate themselves as brightenings of peaks of line profile (k2r or k2v peaks of the Mg\,II\,k). They also have characteristic line shifts (redshift of the k2v peak for the downflows, and blueshift of the k2r peak for the upflows). Thus, because the Mg\,II lines are very sensitive to the upflows and downflows in the upper chromosphere, their spectrograms may serve as good indicators of such events.

\section{Summary}

We presented results of the non-LTE modeling of the H$\alpha$ and Mg\,II\,h\&k lines. Spectral observation of these lines and interpretation of their profiles are very important for understanding the dynamic processes in the solar chromosphere. The H$\alpha$ line is observed with the VIS instrument at NST, and Mg\,II\,h\&k line profiles are now obtained by the NASA's IRIS spacecraft. We described our current progress in the implementation of codes for the non-LTE modeling, and explained the selection of the RH code \cite{Uitenbroek01} as the most appropriate for our research. From the available Hydrogen and Magnesium atomic models we selected the most suitable for our modeling. We considered evolution of atmospheric properties in a selected region of the radiative MHD simulations of the solar atmosphere obtained with the Bifrost code \cite{Gudiksen11}, and calculated synthetic H$\alpha$ and Mg\,II\,h\&k line profiles. We also calculated the oscillation power spectra for the line profile variations as a function of wavelength, and compared with the power spectra of the atmospheric properties, calculated as a function of height.

Three points were selected for each line: the line continuum point, the near wing point and the line center point. We found correlations between the power spectra of intensities at these line points and the atmospheric power spectra at the heights corresponding to the average optical ${\tau}=1$ depth for these line points. The spectra of near wings and line core reveal excitation of high-frequency oscillations in the upper chromosphere, presumably due to non-linear effects and shocks. We also considered upflows and downflows in this data set and discussed possibility of their detection in spectrograms. Further work is needed to study the correlations between the atmospheric parameters and the line profiles.

\section{Acknowledgements}

We thank Han Uitenbroek and Tiago Pereira for their support during working with the RH code. We also thank the team of Bifrost code developers for availability of their results and an opportunity to work with these data sets. Authors acknowledge the BBSO observing and technical team, the IRIS mission team, the Stanford Solar Group and the NASA Ames Research Center for their contribution and support. The work was partially supported by NASA grants NNX14AB68G and NNX14AB70G.

%\vskip+1.5cm

%\appendix
\begin{appendices}

\section{Notes On Radiative Transfer Modeling}

\subsection{Introduction}

The main aim of this appendix is to present theoretical aspects of radiative transfer methodology and computational approaches used in the RH program. The information presented in this appendix is integrated from papers \cite{Rybicki91}, \cite{Rybicki92}, \cite{Uitenbroek01}. Also some information from the Radiative Transfer lecture slides of Han Uitenbroek (\url{http://folk.uio.no/ada/school_2010/Site/School_Materials.html}) is used here.

\subsection{Theoretical aspects of the problem}

In general the radiative transfer (RT) is a problem of computation of radiation field in gases and plasma for a given distribution of physical parameters, like density, velocity, temperature, chemical composition etc. For our tasks, this is a problem of computation of spectral line profiles formed in the solar atmosphere for the dynamical structure of this atmosphere.

The main RT equation for a plain-parallel atmosphere is:

\vskip-.6cm
\begin{gather}
{\mu}\frac{d I_{{\mu},{\nu}}}{d z} = -\chi _{{\mu},{\nu}}I_{{\mu},{\nu}}+\eta _{{\mu},{\nu}}
\end{gather}

Here $\chi _{{\mu},{\nu}}$ and $\eta _{{\mu},{\nu}}$ are the opacity and emissivity coefficients, $I_{{\mu},{\nu}}$ is intensity, and $\mu$ and $\nu$ are cosine of the angle to the vertical direction, and the radiation frequency respectively. An additional scattering term may be added if the atmosphere has some aerosols or drops. We can rewrite this equation in terms of the source function, $S_{{\mu},{\nu}}$, and optical depth, $\tau _{\nu}$:

\vskip-.6cm
\begin{gather}
d\tau _{\nu}(z) = -\chi _{{\mu},{\nu}}(z)dz  \\
S_{{\mu},{\nu}} = \frac{\eta _{{\mu},{\nu}}}{\chi _{{\mu},{\nu}}}  \\
{\mu}\frac{d I_{{\mu},{\nu}}}{d \tau} = S _{{\mu},{\nu}}-I_{{\mu},{\nu}}
\end{gather}

If the source function, $S _{{\mu},{\nu}}$, is given, ordinary one can obtain a formal solution for this differential equation for specified boundary conditions. This solution is usually expressed in terms of an operator ($\Lambda$-operator) acting on the source function:

\vskip-.6cm
\begin{gather}
I_{{\mu},{\nu}}(\tau) = I _{B_{{\mu},{\nu}}} e^{-\frac{\tau _{B} - \tau}{\mu}} + \frac{1}{\mu}\int_{\tau}^{\tau _{B}} S_{{\mu},{\nu}}(t)e^{-(t-\tau)} dt = \Lambda [S_{{\mu},{\nu}}]
\end{gather}

where $\tau _{B}$ is an optical depth corresponding to the lower bound, and $I _{B_{{\mu},{\nu}}}$ is an intensity at this bound (boundary conditions).

If some overlapping lines (transitions) are needed to be resolved, the $\Lambda$-operator approach becomes inconvenient. The source function of overlapping lines is not an additive value. However, the emissivity coefficient is an additive property and represents a sum of emissivities of overlapping transitions. Thus, it becomes more convenient to express equations in terms of $\Psi$-operator of the emissivity:

\vskip-.6cm
\begin{gather}
\label{eq:Psi-def-6}
I  _{{\mu},{\nu}} = \Psi _{{\mu},{\nu}}[\eta _{{\mu},{\nu}}] = \Lambda _{{\mu},{\nu}} \left[ \frac{\eta _{{\mu},{\nu}}}{\chi _{{\mu},{\nu}}} \right]
\end{gather}

In contrast to the source function, the emissivity is additive parameter. Because of this, the $\Psi$-operator is additive, and can include the background radiation, and overlapping frequency transitions.

The second important equation of the RT problem describes populations of different atomic levels $n _{l}$. In general case, it may be written in the form:

\vskip-.6cm
\begin{gather}
n_{l}\sum\limits_{l'}^{} (R_{ll'}+C_{ll'}) - \sum\limits_{l'}^{}n_{l'}(R_{l'l}+C_{l'l}) = \frac{d n_{l}}{d t}
\end{gather}

Here $R_{ll'}$ are the radiative rate coefficients (which can be described in terms of Einstein's coefficients for bound-bound transitions), and $C_{ll'}$ are the collisional rate coefficients. We assume that the radiation processes are much faster than macroscopic processes and consider the radiation transfer equations in a NTLE statistical equilibrium (SE) approximation:

\vskip-.6cm
\begin{gather}
\label{eq:staeq-main}
n_{l}\sum\limits_{l'}^{} (R_{ll'}+C_{ll'}) - \sum\limits_{l'}^{}n_{l'}(R_{l'l}+C_{l'l}) = 0
\end{gather}

\subsection{NTLE radiation transfer in the CRD approximation}

\subsubsection{Local operator approximation without background continuum}

\emph{The Complete Redistribution function (CRD) approximation assumes the emission profile of a line to be equal to its absorption profile.} In the CRD approximation without influence of background continuum, the RT problem is solved via an iterative scheme for the $\Lambda$-operator. The radiative transfer in lines is characterized by strong absorption in the line core (as well as by strong emission) compared to continuum opacities and emissivities. The presence of the background opacities makes equations more complicated, but does not change the formalism. If we consider only the lines the opacities and emissivities can be written as follows:

\vskip-.6cm
\begin{gather}
\eta _{ll'} (\mu,\nu) = \frac{h\nu}{4\pi}n_{l}A_{ll'}\phi _{ll'}(\mu,\nu)  \\
\chi _{ll'} (\mu,\nu) = \frac{h\nu}{4\pi}(n_{l'}B_{l'l} - n_{l}B_{ll'})\phi _{ll'}(\mu,\nu)
\end{gather}

where $l$ and $l'$ are atomic levels (we assume that if $l>l'$ then $E>E'$ ); $A_{ll'}, B_{l'l}, B_{ll'}$ are the Einstein's spontaneous and stimulated emission coefficients; $n_{l}$ and $n_{l'}$ are the populations of the $l$ and $l'$ levels respectively; $\phi _{ll'}(\mu,\nu)$ is the line absorption profile (the emission profile is considered to be the same). The absorption profile can be written as follows:

\vskip-.6cm
\begin{gather}
\phi _{ll'}(\mu,\nu) = \widetilde{\phi _{ll'}} \left( \nu - \nu _{ll'} - \nu _{ll'} \mu \frac{v(z)}{c} \right)
\end{gather}

where the $\widetilde{\phi _{ll'}}$ is a standard normalized profile for given physical properties of plasma (this may be Gaussian, Lorentz, Voigt or some other profile). When calculating the line profiles in the solar atmosphere it is necessary to take into account velocities $v(z)$ along the line of sight. The line source function in this case has a very simple form:

\vskip-.6cm
\begin{gather}
\label{eq:SF}
S_{ll'}=\frac{n_{l}A_{ll'}}{n_{l'}B_{l'l}-n_{l}B_{ll'}}
\end{gather}

Let us first consider the SE Eq.\,(\ref{eq:staeq-main}) for the bound-bound transitions. Un this case the radiative rate coefficients can be written as follows:

\vskip-.6cm
\begin{gather}
R_{ll'}=A_{ll'}+B_{ll'}\overline{J_{ll'}},\; l>l'  \nonumber \\
R_{ll'}=B_{ll'}\overline{J_{ll'}},\; l<l'
\end{gather}

Here, we introduce mean radiation field $\overline{J_{ll'}}$ integrated over the angle and the line profile: $\overline{J_{ll'}} = \frac{1}{4\pi}\int d\Omega \int d\nu \phi _{ll'} ({\mu},{\nu}) I _{{\mu},{\nu}}$. We can rewrite Eq.\,(\ref{eq:staeq-main}) as:

\vskip-.6cm
\begin{gather}
\sum\limits_{l'<l}^{} [n_{l}A_{ll'} - (n_{l'}B_{l'l} - n_{l}B_{ll'})\overline{J_{ll'}}] - \sum\limits_{l'>l}^{} [n_{l'}A_{l'l} - (n_{l}B_{ll'} - n_{l'}B_{l'l})\overline{J_{ll'}}] + \nonumber \\
\label{eq:SE-with-J-14}
+ \sum\limits_{l'}^{} [n_{l}C_{ll'} -n_{l'}C_{l'l}] = 0
\end{gather}

Equations~(\ref{eq:SF}-\ref{eq:SE-with-J-14}) allow us to calculate the source function $S _{{\mu},{\nu}}$ for intensity $I _{{\mu},{\nu}}$ which, in turn, is determined through the $\Lambda$-operator equation:

\vskip-.6cm
\begin{gather}
I_{\mu,\nu} = \Lambda _{\mu,\nu} [S_{\mu,\nu}]
\end{gather}

Now we need to construct an iterative scheme. The computational costs of the complete $\Lambda$-operator may be very high, and some approximation of the $\Lambda$-operator can be introduced. If we consider $\Lambda ^{*}_{\mu,\nu}$ as an approximation to the exact $\Lambda$-operator, we can write the iterative scheme in this form:

\vskip-.6cm
\begin{gather}
\label{eq:lambda-iteration}
I _{\mu,\nu} = \Lambda ^{*}_{\mu,\nu} [S_{\mu,\nu}] + (\Lambda _{\mu,\nu} - \Lambda ^{*}_{\mu,\nu}) [S^{+}_{\mu,\nu}]
\end{gather}

The $^{+}$ symbol means that the value from the previous iteration step is used. This is the basis of the iterative technique. Here the $S^{+}_{\mu,\nu}$ is the source function at the previous iteration step. For the converged solution: $S^{+}=S$, and, thus, $I_{\mu,\nu} = \Lambda _{\mu,\nu} [S _{\mu,\nu}]$. Using the fact that $\Lambda _{\mu,\nu} [S^{+}_{\mu,\nu}] = I^{+}_{\mu,\nu}$, and introducing discrepancy: $I^{eff}_{\mu,\nu} = I^{+}_{\mu,\nu} - \Lambda ^{*}_{\mu,\nu} [S^{+}_{\mu,\nu}]$, we can rewrite the previous equation:

\vskip-.6cm
\begin{gather}
I _{\mu,\nu} = \Lambda ^{*}_{\mu,\nu} [S_{\mu,\nu}] + I^{eff}_{\mu,\nu}
\end{gather}

Note that $I^{eff}_{\mu,\nu}$ is a function defined at the previous iteration, so it is known. The general algorithm to solve the radiation transfer problem is the following. First, we define an initial solution for the radiation transfer problem: for example, assuming the LTE populations of atomic states at all layers. After this, we calculate emissivity and opacity for each transition. Then, we calculate the source function $S _{{\mu},{\nu}}$ and intensity $I _{{\mu},{\nu}}$ which are functions of frequency and direction. After this we calculate the mean intensity, $\overline{J_{ll'}}$, integrated over the line profile. Note, that at this step it is important to include line-of-sight velocities. After this step, the calculated mean intensity is used in Eq.\,(\ref{eq:staeq-main}) to obtain new populations. Then, the iteration procedure is represented until the discrepancy becomes sufficiently small. This scheme is used in some form in all our cases.

We use a local approximation for construction of the simplified operator, $\Lambda ^{*}$. This means that $\Lambda ^{*}$ in a particular layer depends only on the source function of the same layer. Thus, the local approximation means that the diagonal part of the original operator is used as $\Lambda ^{*}$. The $\overline{J_{ll'}}$ radiation field term is calculated as:

\vskip-.6cm
\begin{gather}
\overline{\Lambda _{ll'}^{*}} = \int d\Omega \int d\nu \phi _{ll'} ({\mu},{\nu}) \Lambda ^{*} _{{\mu},{\nu}} \\
\overline{J ^{eff}_{ll\prime}} = \int d\Omega \int d\nu \phi _{ll\prime} ({\mu},{\nu}) I ^{eff} _{{\mu},{\nu}} \\
\overline{J _{ll\prime}} = \overline{\Lambda _{ll\prime}^{*}} S_{ll\prime} + \overline{J ^{eff}_{ll\prime}}
\end{gather}

Using this result, and also the source function Eq.\,(\ref{eq:SF}) in the statistical equilibrium equation (\ref{eq:SE-with-J-14}), we obtain:

\vskip-.6cm
\begin{gather}
\sum\limits_{l'<l}^{} [n_{l}A_{ll'}(1 - \overline{\Lambda _{ll'}^{*}}) - (n_{l'}B_{l'l} - n_{l}B_{ll'})\overline{J^{eff}_{ll'}}] - \nonumber \\
- \sum\limits_{l'>l}^{} [n_{l'}A_{l'l}(1 - \overline{\Lambda _{ll'}^{*}}) - (n_{l}B_{ll'} - n_{l'}B_{l'l})\overline{J^{eff}_{ll'}}] + \nonumber \\
+ \sum\limits_{l'}^{} [n_{l}C_{ll'} -n_{l'}C_{l'l}] = 0
\end{gather}

This linear system for the populations, $n_{l}$, is solved by a standard LU technique.

\subsubsection{Local operator approximation with background continuum}

To take into account the background continuum we include additional terms for opacities and emissivities:

\vskip-.6cm
\begin{gather}
\eta _{\mu,\nu} = \eta _{ll'}({\mu},{\nu}) + \eta _{cll'} \\
\chi _{\mu,\nu} = \chi _{ll'}({\mu},{\nu}) + \chi _{cll'}
\end{gather}

where $\eta _{cll'}$ and $\chi _{cll'}$ are the background emissivity and opacity corresponding to the line frequency, $\nu _{ll'}$. Then, the source function has the form:

\vskip-.6cm
\begin{gather}
\label{eq:sf-linear-24}
S _{\mu,\nu} = \frac{\eta _{\mu,\nu}}{\chi _{\mu,\nu}} = r_{ll'}S_{ll'} + (1 - r_{ll'})S_{cll'},
\end{gather}

where the $S_{cll'} = \frac{\eta _{cll'}}{\chi _{cll'}}$ and $r_{ll'} = \frac{\chi _{ll'}}{\chi _{ll'} + \chi _{cll'}}$. Note that the line coefficients $\chi _{ll'} = \chi _{ll'}(\mu,\nu)$ and $\eta _{ll'} = \eta _{ll'}(\mu,\nu)$ are functions of $\mu$ and $\nu$, while all the continuum coefficients $\eta _{cll'}$, $\chi _{cll'}$ are constants.

This representation of the source function leads to nonlinear statistical equilibrium equations. However, it is possible to simplify this by making preconditioning, and approximating the source function as:

\vskip-.6cm
\begin{gather}
S _{\mu,\nu} = r_{ll'}^{+}S_{ll'} + (1 - r_{ll'}^{+})S_{cll'}, \\
r_{ll'}^{+} = \frac{\chi _{ll'}^{+}}{\chi _{ll'}^{+} + \chi _{cll'}}
\end{gather}

Where the $r_{ll'}^{+}$ parameter is determined from the previous iteration step values. This approach leads to linear statistical equilibrium equations. Here we follow authors of paper \cite{Rybicki91}, who argue that this substitution does not affect the final solution if the iteration procedure converges.

Now, from Eq.\,(\ref{eq:lambda-iteration}) we obtain the intensity:

\vskip-.6cm
\begin{gather}
I _{\mu,\nu} = \Lambda ^{*}_{\mu,\nu}r_{ll'}^{+}S_{ll'} + \widetilde{I^{eff}_{\mu,\nu}}, \\
\widetilde{I^{eff}_{\mu,\nu}} = \Lambda _{\mu,\nu}[S^{+}_{\mu,\nu}] - \Lambda _{\mu,\nu}^{*}r_{ll'}^{+}S_{ll'}^{+} = I_{\mu,\nu}^{+} - \Lambda _{\mu,\nu}^{*}r_{ll'}^{+}S_{ll'}^{+}
\end{gather}

The continuum source function is included here in the value of $S_{\mu,\nu}^{+}$. Of course, we need to use the same integration procedure as in the previous derivation:

\vskip-.6cm
\begin{gather}
\widetilde{\Lambda _{ll'}^{*}} = \int d\Omega \int d\nu \phi _{ll'} ({\mu},{\nu}) \Lambda ^{*} _{{\mu},{\nu}}r_{ll'}^{+} \\
\widetilde{J ^{eff}_{ll\prime}} = \int d\Omega \int d\nu \phi _{ll\prime} ({\mu},{\nu}) I ^{eff} _{{\mu},{\nu}} \\
\overline{J _{ll\prime}} = \widetilde{\Lambda _{ll\prime}^{*}} S_{ll\prime} + \widetilde{J ^{eff}_{ll\prime}}
\end{gather}

Substituting these expressions in Eq.\,(\ref{eq:SE-with-J-14}) (SE equation), we have:

\vskip-.6cm
\begin{gather}
\sum\limits_{l'<l}^{} [n_{l}A_{ll'}(1 - \widetilde{\Lambda _{ll'}^{*}}) - (n_{l'}B_{l'l} - n_{l}B_{ll'})\widetilde{J^{eff}_{ll'}}] - \nonumber \\
- \sum\limits_{l'>l}^{} [n_{l'}A_{l'l}(1 - \widetilde{\Lambda _{ll'}^{*}}) - (n_{l}B_{ll'} - n_{l'}B_{l'l})\widetilde{J^{eff}_{ll'}}] + \nonumber \\
+ \sum\limits_{l'}^{} [n_{l}C_{ll'} -n_{l'}C_{l'l}] = 0
\end{gather}

This linear system is easily solved numerically. Note that the local approximation replaces the $\Lambda$-operator with multiplication of its diagonal elements.

\subsubsection{Nonlocal operator approximation with background continuum}

The non-locality of the $\Lambda$-operator means that the matrix representing the operator is no longer diagonal. This section shows that due to the non-locality the final SE equations are nonlinear.

Rybicki and Hummer \cite{Rybicki91} introduced the following representation of the source function:

\vskip-.6cm
\begin{gather}
\label{eq:SF-linear-33}
S _{\mu,\nu} = r_{ll'}^{*}S_{ll'} + (1 - r_{ll'}^{*})S_{cll'}, \\
r_{ll'}^{*} = \frac{\chi _{ll'}}{\chi _{ll'}^{+} + \chi _{cll'}}
\end{gather}

where $ r_{ll'}^{*}$  differs from $ r_{ll'}^{+}$ because the numerator is not taken from the previous iteration but unknown. In this case, the intensity obtained from Eq.\,(\ref{eq:SF}) is:

\vskip-.6cm
\begin{gather}
I_{\mu,\nu} = \Lambda ^{*}_{\mu,\nu} [r^{*}_{ll'}S_{ll'}] + \widetilde{I^{eff}_{\mu,\nu}}, \\
\widetilde{I^{eff}_{\mu,\nu}} = \Lambda _{\mu,\nu} [S^{+}_{\mu,\nu}] - \Lambda _{\mu,\nu}^{*}[r^{*}_{ll'}S^{+}_{\mu,\nu}] = I^{+}_{\mu,\nu} - \Lambda _{\mu,\nu}^{*}[r^{*+}_{ll'}S^{+}_{\mu,\nu}]
\end{gather}

The mean profile intensity can be expressed now in this form:

\begin{gather}
\widetilde{J _{ll\prime}} = \widetilde{\Lambda _{ll\prime}^{*}} [S_{ll\prime}] + \widetilde{J ^{eff}_{ll\prime}}, \\
\widetilde{\Lambda _{ll'}^{*}}[...] = \int d\Omega \int d\nu \phi _{ll'} ({\mu},{\nu}) \Lambda ^{*} _{{\mu},{\nu}}[r_{ll'}^{*} ...], \\
\widetilde{J ^{eff}_{ll\prime}} = \int d\Omega \int d\nu \phi _{ll\prime} ({\mu},{\nu}) \widetilde{I ^{eff} _{{\mu},{\nu}}}
\end{gather}

Now, we substitute these in the SE equation~(\ref{eq:SE-with-J-14}), and obtain:

\vskip-.6cm
\begin{gather}
\sum\limits_{l'<l}^{} \left[ n_{l}A_{ll'} - (n_{l'}B_{l'l} - n_{l}B_{ll'})\widetilde{\Lambda _{ll'}^{*}}[S_{ll'}] - (n_{l'}B_{l'l} - n_{l}B_{ll'})\widetilde{J^{eff}_{ll'}} \right] - \nonumber \\
- \sum\limits_{l'>l}^{} \left[ n_{l'}A_{l'l} - (n_{l}B_{ll'} - n_{l'}B_{l'l})\widetilde{\Lambda _{ll'}^{*}}[S_{ll'}] - (n_{l}B_{ll'} - n_{l'}B_{l'l})\widetilde{J^{eff}_{ll'}} \right] + \nonumber \\
+ \sum\limits_{l'}^{} \left[ n_{l}C_{ll'} -n_{l'}C_{l'l} \right] = 0
\end{gather}

The $\widetilde{\Lambda _{ll'}^{*}}[S_{ll'}]$ operator can be derived from the previous equations in the following form:

\vskip-.6cm
\begin{gather}
\widetilde{\Lambda _{ll'}^{*}}[S_{ll'}] = A_{ll'}\frac{h\nu}{4\pi}\int d\Omega \int d\nu \phi _{ll'} \Lambda ^{*}_{\mu,\nu} \left[ \frac{\phi _{ll'}}{\chi ^{+}_{ll'} + \chi _{cll'} }n_{l} \right]
\end{gather}

It also linearly depends on the populations, $n_{l}$. This forms nonlinear terms in the SE equation: $(n_{l'}B_{l'l} - n_{l}B_{ll'})\widetilde{\Lambda _{ll'}^{*}}[S_{ll'}]$ and $(n_{l}B_{ll'} - n_{l'}B_{l'l})\widetilde{\Lambda _{ll'}^{*}}[S_{ll'}]$. To solve this we apply preconditioning: by using the populations as ones from the previous iteration step in one of terms: $(n_{l}B_{ll'} - n_{l'}B_{l'l}) \rightarrow (n_{l}^{+}B_{ll'} - n_{l'}^{+}B_{l'l})$ and $(n_{l'}B_{l'l} - n_{l}B_{ll'}) \rightarrow (n_{l'}^{+}B_{l'l} - n_{l}^{+}B_{ll'})$. Now we get the statistical equilibrium equation in the form of a linear system:

\vskip-.6cm
\begin{gather}
\sum\limits_{l'<l}^{} \left[ n_{l}A_{ll'} - (n_{l'}^{+}B_{l'l} - n_{l}^{+}B_{ll'})\widetilde{\Lambda _{ll'}^{*}}[S_{ll'}] - (n_{l'}B_{l'l} - n_{l}B_{ll'})\widetilde{J^{eff}_{ll'}} \right] - \nonumber \\
- \sum\limits_{l'>l}^{} \left[ n_{l'}A_{l'l} - (n_{l}^{+}B_{ll'} - n_{l'}^{+}B_{l'l})\widetilde{\Lambda _{ll'}^{*}}[S_{ll'}] - (n_{l}B_{ll'} - n_{l'}B_{l'l})\widetilde{J^{eff}_{ll'}} \right] + \nonumber \\
+ \sum\limits_{l'}^{} \left[ n_{l}C_{ll'} -n_{l'}C_{l'l} \right] = 0
\end{gather}

\subsubsection{$\Psi$-operator approach for the overlapping transitions}

The second paper of Rybicki and Hummer \cite{Rybicki92} involves $\Psi$-operator (which is determined according to Eq.\,(\ref{eq:Psi-def-6}) and discussed  in Sec.\,2) to solve the SE and RT equations for the atomic transitions overlapping in frequency. The reason for this is that the $\Psi$-operator acts on emissivity which is an additive property and represents a sum of emissivities of different transitions and background continuum emissivity. In Eq.\,(\ref{eq:sf-linear-24}) and~(\ref{eq:SF-linear-33}) the source function is represented as a linear superposition of a background source function and a line source function. However, in the general case, with many overlapping transitions it is impractical to represent the total source function as a linear combination of source functions of each additional transition.

We introduce some new parameters and expressions that are needed for construction of the iterative scheme. The SE equation can be written as follows:

\vskip-.6cm
\begin{gather}
\sum\limits_{l'}^{} n_{l'}(C_{l'l} + R_{l'l}) = n_{l}\sum\limits_{l'}^{} (C_{l'l} + R_{l'l})
\end{gather}

If the Einstein coefficients are known, and the line profile of each layer is defined (usually as a Gaussian or Voigt profile with the Doppler shift determined by the large-scale line-of-sight velocities), we can introduce two new quantities for every additional transition $ll'$:

\vskip-.6cm
\begin{gather}
\label{eq:Ucoef-CRD-44}
U_{ll'}(\mu,\nu) = \frac{h\nu}{4\pi}A_{ll'}\phi _{ll'} (\mu,\nu),\; l>l', \\
U_{ll'}(\mu,\nu) = 0,\; l<l', \\
\label{eq:Vcoef-CRD-46}
V_{ll'}(\mu,\nu) = \frac{h\nu}{4\pi}B_{ll'}\phi _{ll'} (\mu,\nu)
\end{gather}

The opacities and emissivities can now be expressed in terms of these new parameters. For the relationship between the populations of the $l$ and $l'$ levels we can write:

\vskip-.6cm
\begin{gather}
\chi _{ll'} = n _{l'}V_{l'l} - n_{l}V_{ll'} \\
\eta _{ll'} = n_{l}U_{ll'}
\end{gather}

To determine the total opacities and emissivities as functions of frequency $\nu$ and direction $\mu$ we can sum over all the transitions. At this point the overlapping is taken into account:

\vskip-.6cm
\begin{gather}
\chi _{\mu,\nu} = \sum\limits_{l>l'}^{} \chi _{ll'} + \chi _{c} = \sum\limits_{l>l'}^{} n _{l'}V_{l'l} - n_{l}V_{ll'} + \chi _{c} \\
\label{eq:tot-op-50}
\eta _{\mu,\nu} = \sum\limits_{ll'}^{} \eta _{ll'} + \eta _{c} = \sum\limits_{ll'}^{} n_{l}U_{ll'} + \eta _{c}
\end{gather}

Here $\chi _{c}$ and $\eta _{c}$ are background opacity and emissivity coefficients corresponding to a particular line frequency, and assumed constant along the line profile. The radiative rate coefficient for transitions between $l$ and $l'$ levels can be determined as follows:

\vskip-.6cm
\begin{gather}
R_{ll'} = \int d\Omega \int \frac{d\nu}{h\nu} \left[ U_{ll'}(\mu,\nu) + V_{ll'}(\mu,\nu)I_{\mu,\nu} \right]
\end{gather}

and the SE equation can be written in a self-consistent form:

\vskip-.6cm
\begin{gather}
\sum\limits_{l'}^{} n_{l'}C_{l'l} + \sum\limits_{l'}^{}\int d\Omega \int \frac{d\nu}{h\nu} \left[ n_{l'}U_{l'l}(\mu,\nu) + n_{l'}V_{l'l}(\mu,\nu)I_{\mu,\nu} \right] = \nonumber \\
\label{eq:RT-full-52}
= \sum\limits_{l'}^{} n_{l}C_{ll'} + \sum\limits_{l'}^{}\int d\Omega \int \frac{d\nu}{h\nu} \left[ n_{l}U_{ll'}(\mu,\nu) + n_{l}V_{ll'}(\mu,\nu)I_{\mu,\nu} \right]
\end{gather}

Consider numerical solutions for this case. The iteration procedure for the $\Psi$-operator:

\vskip-.6cm
\begin{gather}
I_{\mu,\nu} = \Psi _{\mu,\nu}[\eta _{\mu,\nu}]
\end{gather}

can be formulated in the same way as for the $\Lambda$-operator:

\vskip-.6cm
\begin{gather}
\label{eq:iter-psi-54}
I_{\mu,\nu} = \Psi ^{*}_{\mu,\nu}[\eta _{\mu,\nu}] + (\Psi _{\mu,\nu} - \Psi ^{*}_{\mu,\nu})[\eta _{\mu,\nu}^{+}], \\
\eta _{\mu,\nu}^{+} = \sum\limits_{ll'}^{} n^{+}_{l}U_{ll'} + \eta _{c}
\end{gather}

The $n^{+}_{l}$ is the population of the $l$ level at the previous iteration step, and the $\Psi ^{*}_{\mu,\nu}$ is an approximate operator. The RH radiative transfer program uses the local approximation for this operator \cite{Uitenbroek01}. The approximation may be more complex, and here we consider a general case. Because the $\Psi$-operator also uses opacities when deriving the intensity, it is necessary to use their values from the previous iterations.

Substituting emissivity from Eq.\,(\ref{eq:tot-op-50}) to Eq.\,(\ref{eq:iter-psi-54}), we get:

\vskip-.6cm
\begin{gather}
\label{eq:iter-psi-56}
I_{\mu,\nu} = \Psi _{\mu,\nu} [\eta ^{+}_{\mu,\nu}] - \sum\limits_{mm'}^{} \Psi ^{*}_{\mu,\nu} [n^{+}_{m} U_{mm'}] + \sum\limits_{mm'}^{} \Psi ^{*}_{\mu,\nu} [n_{m} U_{mm'}]
\end{gather}

The background opacity is assumed to be the same for every iteration, and does not appear in this equation. The SE equation can be obtained by substituting Eq.\,(\ref{eq:iter-psi-56}) in Eq.\,(\ref{eq:RT-full-52}):

\vskip-.6cm
\begin{gather}
\sum\limits_{l'}^{} n_{l'}C_{l'l} + \sum\limits_{l'}^{}\int d\Omega \int \frac{d\nu}{h\nu} (n_{l'}U_{l'l} + n_{l'}V_{l'l}\Psi _{\mu,\nu}[\eta _{\mu,\nu}^{+}] - \nonumber \\
- \sum\limits_{mm'}^{} n_{l'}V_{l'l}\Psi ^{*}_{\mu,\nu}[n^{+}_{m} U_{mm'}] + \sum\limits_{mm'}^{}n_{l'}V_{l'l}\Psi ^{*}_{\mu,\nu}[n_{m}U_{mm'}]) = \nonumber \\
= \sum\limits_{l'}^{} n_{l}C_{ll'} + \sum\limits_{l'}^{}\int d\Omega \int \frac{d\nu}{h\nu} (n_{l}U_{ll'} + n_{l}V_{ll'}\Psi _{\mu,\nu}[\eta _{\mu,\nu}^{+}] - \nonumber \\
\label{eq:57-stat}
- \sum\limits_{mm'}^{} n_{l}V_{ll'}\Psi ^{*}_{\mu,\nu}[n^{+}_{m} U_{mm'}] + \sum\limits_{mm'}^{}n_{l}V_{ll'}\Psi ^{*}_{\mu,\nu}[n_{m}U_{mm'}])
\end{gather}

There are two nonlinear terms which need preconditioning in this equation:\\
$\sum\limits_{mm'}^{}n_{l'}V_{l'l}\Psi ^{*}_{\mu,\nu}[n_{m}U_{mm'}]$ and $\sum\limits_{mm'}^{}n_{l}V_{ll'}\Psi ^{*}_{\mu,\nu}[n_{m}U_{mm'}]$. The problem here is that if we substitute the populations $n_{m}$ with their previous values $n_{m}^{+}$, the last two terms in Eq.\,(\ref{eq:57-stat}) cancel. Authors \cite{Rybicki92} claim that if this approach is adapted, then the iteration procedure represents the classical $\Lambda$-iteration scheme. However, we know that the classical $\Lambda$-iterations scheme does not deal with overlapping transition. Thus, this preconditioning strategy fails. It is necessary to obtain some of $n_{l}$ values from the previous iteration step.

For instance, we still can use some of the $n_{m}$ values (but not all of them) from the previous solution. This approach leads to two basic strategies.

1. Full preconditioning strategy. According to this strategy, all the $n_{l}$ values in nonlinear terms are replaced with the values from the previous iteration step. In this case, we obtain an equation:

\vskip-.6cm
\begin{gather}
\sum\limits_{l'}^{} n_{l'}C_{l'l} + \sum\limits_{l'}^{}\int d\Omega \int \frac{d\nu}{h\nu} (n_{l'}U_{l'l} + n_{l'}V_{l'l}\Psi _{\mu,\nu}[\eta _{\mu,\nu}^{+}] - \nonumber \\
- \sum\limits_{mm'}^{} n_{l'}V_{l'l}\Psi ^{*}_{\mu,\nu}[n^{+}_{m} U_{mm'}] + \sum\limits_{mm'}^{}n_{l'}^{+}V_{l'l}\Psi ^{*}_{\mu,\nu}[n_{m}U_{mm'}]) = \nonumber \\
= \sum\limits_{l'}^{} n_{l}C_{ll'} + \sum\limits_{l'}^{}\int d\Omega \int \frac{d\nu}{h\nu} (n_{l}U_{ll'} + n_{l}V_{ll'}\Psi _{\mu,\nu}[\eta _{\mu,\nu}^{+}] - \nonumber \\
- \sum\limits_{mm'}^{} n_{l}V_{ll'}\Psi ^{*}_{\mu,\nu}[n^{+}_{m} U_{mm'}] + \sum\limits_{mm'}^{}n_{l}^{+}V_{ll'}\Psi ^{*}_{\mu,\nu}[n_{m}U_{mm'}])
\end{gather}

However, we are dealing with the sum of $n_{m}$, and it is possible to make preconditions for some of these values also. Authors \cite{Rybicki92} also note that the full preconditioning strategy may lead to coupling between levels that are physically uncoupled. Thus, they introduce another approach:

2. Preconditioning within the same transition only. The authors mention the fact that the overlapping of functions $V_{ll'}$ and $U_{mm'}$ is much greater for cases when indices $mm'$ are the same as $ll'$ or $l'l$ than in other cases. Thus, it is possible to take into account only the terms with $m=l$ and $m=l'$. For other $n_{m}$ it is possible to use values from the previous iteration. This approach leads to cancellations of terms including $n_{m}^{+}$, and the scheme in this case is:

\vskip-.6cm
\begin{gather}
\sum\limits_{l'}^{} n_{l'}C_{l'l} + \sum\limits_{l'}^{}\int d\Omega \int \frac{d\nu}{h\nu} (n_{l'}U_{l'l} + n_{l'}V_{l'l}\Psi _{\mu,\nu}[\eta _{\mu,\nu}^{+}] - \nonumber \\
- n_{l'}V_{l'l}\Psi ^{*}_{\mu,\nu}[n^{+}_{l} U_{ll'}] - n_{l'}V_{l'l}\Psi ^{*}_{\mu,\nu}[n^{+}_{l'} U_{l'l}] + n_{l'}^{+}V_{l'l}\Psi ^{*}_{\mu,\nu}[n_{l} U_{ll'}] + n_{l'}^{+}V_{l'l}\Psi ^{*}_{\mu,\nu}[n_{l'} U_{l'l}] = \nonumber \\
= \sum\limits_{l'}^{} n_{l}C_{ll'} + \sum\limits_{l'}^{}\int d\Omega \int \frac{d\nu}{h\nu} (n_{l}U_{ll'} + n_{l}V_{ll'}\Psi _{\mu,\nu}[\eta _{\mu,\nu}^{+}] - \nonumber \\
- n_{l}V_{ll'}\Psi ^{*}_{\mu,\nu}[n^{+}_{l} U_{ll'}] - n_{l}V_{ll'}\Psi ^{*}_{\mu,\nu}[n^{+}_{l'} U_{l'l}] + n_{l}^{+}V_{ll'}\Psi ^{*}_{\mu,\nu}[n_{l} U_{ll'}] + n_{l}^{+}V_{ll'}\Psi ^{*}_{\mu,\nu}[n_{l'} U_{l'l}]
\end{gather}

\subsection{$\Psi$-operator approach for the PRD case}

The Partial Redistribution function (PRD) is an approach when the emission profile of an atom is not independent from the direction and frequency of an absorbed photon. In this case the emission profile is determined by the absorption profile, radiation field, populations of atomic layers, direction etc. The PRD effects are not so important for the H$\alpha$ line profile (Figure 5 of Leenaarts et al \cite{Leenaarts12}) . However, these are very important for the modeling of Mg~II lines (Figure 11 of Leenaarts et al \cite{Leenaarts13a}). Because one of our aims is calculation of Mg~II lines, we need to take this effect into account. It is possible to simplify the PRD approach, and consider the dependence of the emission profile only on frequency of absorbed photon, neglecting the dependence on the direction. In this case PRD approach is called angle-independent PRD.

Because the emission profile depends on both, the populations and radiation field, an additional iteration loop is required for computations. This description follows the paper of Uitenbroek \cite{Uitenbroek01}.

As one can see from the Eq.\,(\ref{eq:Ucoef-CRD-44}-\ref{eq:Vcoef-CRD-46}), the emission profile appears in the transition coefficients $U$ and $V$. If the $\phi _{ij}(\nu,\vec{n})$ is an absorption profile function of the $ij$ transition, which depends on the frequency $\nu$ and direction $\vec{n}$, and $\psi _{ij}(\nu,\vec{n})$ is an emission profile function, then we can rewrite the emission and absorption coefficients as follows:

\vskip-.6cm
\begin{gather}
\chi _{ij}(\nu,\vec{n}) = n _{i}V_{ij}(\nu,\vec{n}) - n_{j}V_{ji}(\nu,\vec{n}) \\
\eta _{ij}(\nu,\vec{n}) = n_{j}U_{ji}(\nu,\vec{n})
\end{gather}

where the functions $U_{ij}(\nu,\vec{n})$ and $n_{j}V_{ji}(\nu,\vec{n})$ are defined in terms of the Einstein coefficients, and the absorption and emission profiles:

\vskip-.6cm
\begin{gather}
V_{ij} = \frac{h\nu}{4\pi} B_{ij} \phi _{ij}(\nu,\vec{n}), \nonumber \\
V_{ji} = \frac{h\nu}{4\pi} B_{ji} \psi _{ij}(\nu,\vec{n}), \nonumber \\
\label{eq:U-and-V-def}
U_{ji} = \frac{h\nu}{4\pi} A_{ji} \psi _{ij}(\nu,\vec{n})
\end{gather}

In this appendix we consider only bound-bound transitions. However, the formulation is similar if we also include bound-free transitions in terms of $V$ and $U$. See the original paper of Uitenbroek \cite{Uitenbroek01} for details. The total emissivity in direction $\vec{n}$ and frequency $\nu$ can be written as follows:

\vskip-.6cm
\begin{gather}
\chi _{tot}(\nu,\vec{n}) = \chi _{c}(\nu,\vec{n}) + \sum\limits_{i}^{}\sum\limits_{j>i}^{}n_{i}V_{ij}(\nu,\vec{n}) - n_{j}V_{ji}(\nu,\vec{n}), \nonumber \\
\eta _{tot}(\nu,\vec{n}) = \eta _{c}(\nu,\vec{n}) + \sum\limits_{i}^{}\sum\limits_{j>i}^{}n_{j}U_{ji}(\nu,\vec{n})
\end{gather}

The emission profile $\psi _{ij}(\nu,\vec{n})$ dependence on the absorption profile $\phi _{ij}(\nu,\vec{n})$ is described by Uitenbroek \cite{Uitenbroek01}:

\vskip-.6cm
\begin{gather}
\label{eq:R-maindef-64}
\psi _{ij}(\nu,\vec{n})=\phi _{ij}(\nu,\vec{n}) \left(1+\frac{\sum_{k<j}n_{k}B_{kj}}{n_{j}P_{j}}\int \frac{d\Omega}{4\pi} \int d{\nu}'I({\nu}',\vec{n}) \times \right. \nonumber \\
\left. \times \left[ \frac{R_{kji}(\nu,\vec{n};{\nu}',\vec{n'})}{\phi _{ij}(\nu,\vec{n})} - \phi _{kj}({\nu}',\vec{n'}) \right] \right)
\end{gather}

The total depopulation rate, $P_{j}$, of level $j$ is defined as the sum of radiative $R_{jk}$ and collisional $C_{jk}$ rates $P_{j} = \sum\limits_{k \ne j}^{}(C_{jk} + R_{jk})$. The $R_{kji}$ is defined as follows: the $R_{kji}({\nu}',\vec{n'};{\nu},\vec{n})d{\nu}'d\nu \frac{d{\Omega}'}{4\pi} \frac{d\Omega}{4\pi}$ is the probability of photon scattered and absorbed in a solid angle cone $d{\Omega}'$ in $n'$ direction and frequency interval $({\nu}',{\nu}'+\delta {\nu}')$ in line $(k, j)$ to be re-emitted into solid angle $d\Omega$ in direction $\vec{n}$ at with frequency between $({\nu},{\nu}+\delta {\nu})$ in line (i, j). This function is normalized: the integral of this function over the angles and frequencies of this function is unity. This function also satisfies two conditions:

\vskip-.6cm
\begin{gather}
\label{eq:R-cond-1-65}
\frac{1}{4\pi}\int d{\Omega}' \int d{\nu}' R_{kji} ({\nu}',\vec{n'},\nu,\vec{n}) = \phi _{ij} (\nu,\vec{n}) \\
\label{eq:R-cond-2-66}
\frac{1}{4\pi}\int d{\Omega} \int d{\nu} R_{kji} ({\nu}',\vec{n'},\nu,\vec{n}) = \phi _{kj} ({\nu}',\vec{n'})
\end{gather}

Equation~\ref{eq:R-cond-1-65} describes the fact that if the radiation field is uniform and constant in frequency and direction, the emission profile of the $ji$ transition is defined as a "natural" line profile (e.g. determined by velocities, collisions, but independent from direction and frequency of absorbed photon). Equation~(\ref{eq:R-cond-2-66}) means that if we integrate over all the emission lines, we get the natural absorption profile of the $kj$ transition.

Under the two assumptions (first, the lower level broadening in the atom frame is negligible; second, the compete redistribution in the atom frame can be approximated as complete redistribution in the laboratory frame) the redistribution function $R_{ijk}$ may be simplified \cite{Hubeny82}:

\vskip-.6cm
\begin{gather}
R_{kij} = \gamma R^{II}_{kij} + (1 - \gamma) \phi _{kj} \phi _{ij}, \\
\gamma = \frac{P_{j}}{P_{j} + Q_{j}^{E}}
\end{gather}

Here, $Q_{j}^{E}$ is the upper level rate of elastic collisions (usually sum of Van der Waals and Stark impact broadening rates). We can write the ratio of the two profile as a function $\rho = {\psi}/{\phi}$ and write the Eq.\,(\ref{eq:R-maindef-64}) as the following:

\vskip-.6cm
\begin{gather}
\rho _{ij} (\nu,\vec{n}) = 1 + \frac{\gamma \sum_{k<j} n_{k}B_{kj}}{n_{j}P_{j}} \int\frac{d\Omega '}{4\pi} \int d\nu 'I(\nu ',\vec{n} ') \times \nonumber \\
\label{eq:rho-main}
\times \left[ \frac{R^{II}_{kji}(\nu, \vec{n}, \nu ', \vec{n'})}{\phi _{ij} (\nu, \vec{n})} - \phi _{kj} (\nu ', \vec{n'}) \right]
\end{gather}

To derive the statistical equilibrium equation, we can apply all the same procedures of the previous section: introduce an approximate $\Psi$-operator, write an iterative scheme for the intensity, substitute emissivities in terms of $U$ and $V$ functions (Eq.\,(\ref{eq:U-and-V-def})), and substitute the intensity in the SE equation. Finally, we get the following:

\vskip-.6cm
\begin{gather}
\sum\limits_{l'}^{} n_{l'}C_{l'l} + \sum\limits_{l'}^{}\int d\Omega \int \frac{d\nu}{h\nu} (n_{l'}U_{l'l} + n_{l'}V_{l'l}I_{\nu,\vec{n}}^{eff} + n_{l'}V_{l'l}\sum\limits_{j} \sum\limits_{i<j} \Psi ^{*}_{\nu,\vec{n}} [n_{j} U_{ji}]) = \nonumber \\
\sum\limits_{l'}^{} n_{l}C_{ll'} + \sum\limits_{l}^{}\int d\Omega \int \frac{d\nu}{h\nu} (n_{l}U_{ll'} + n_{l}V_{ll'}I_{\nu,\vec{n}}^{eff} + n_{l}V_{ll'}\sum\limits_{j} \sum\limits_{i<j} \Psi ^{*}_{\nu,\vec{n}} [n_{j} U_{ji}]) \\
I^{eff}_{\nu,\vec{n}} = I^{+}_{\nu,\vec{n}} - \Psi ^{*}_{\nu, \vec{n}} \left[ \sum\limits_{j} \sum\limits_{i<j} n^{+}_{j}U^{+}_{ji} \right]
\end{gather}

In the CRD assumption, the nonlinearity of the last two terms in both sides is avoided. However, in the PRD formulation, terms $U$ and $V$ are nonlinear and should be substituted with their values from the previous iteration step, $U^{+}$ and $V^{+}$, respectively. The final form of the RT equation is the following:

\vskip-.6cm
\begin{gather}
\sum\limits_{l'}n_{l'}\Gamma _{l'l} = 0, \\
\Gamma _{l'l} = C_{l'l} + \int d\Omega \int \frac{d\nu}{h\nu} \{ U_{l'l}^{+} + V^{+}_{l'l}I_{\nu,\vec{n}}^{eff} - \sum\limits_{j}(n^{+}_{l}V^{+}_{lj} - n^{+}_{jl}) \sum\limits_{i<l'} \Psi ^{*}_{\nu,\vec{n}} [U^{+}_{l'i}] \} - \nonumber \\
 - \delta _{l'l} \sum\limits_{l''} \{ C_{ll''} + \int d\Omega \int \frac{d\nu}{h\nu} [U^{+}_{ll''} + V^{+}_{ll''}I^{eff}_{\nu,\vec{n}}]  \}, \\
\sum\limits_{j}(n^{+}_{l}V^{+}_{lj} - n^{+}_{jl}) = \sum\limits_{j>l} \chi ^{+}_{lj} - \sum\limits_{j<l} \chi ^{+}_{jl}
\end{gather}

We decided to use $U$ and $V$ parameters from the previous iteration step. However, these parameters which correspond to the emission are unknown because they depend on the radiation field. In fact, all we know from the previous iteration step are the populations. To obtain the radiation field, we need to solve the radiative transfer equation:

\vskip-.6cm
\begin{gather}
\label{eq:RT-ndir}
\vec{n} \nabla I(\nu,\vec{n}) = -\chi ^{tot}I(\nu,\vec{n}) + \frac{h\nu}{4\pi} \sum\limits^{PRD}_{i,j}n_{j}A_{ji}\phi _{ij}\rho _{ij}[I(\nu,\vec{n})] + \sum\limits^{other}_{i,j} n_{j}U_{ji} + \eta ^{c}
\end{gather}

Here, the "PRD" means the sum over all lines for which we take into account the PRD effects, and "other" means the sum over all the transitions except the PRD ones. If we assume that the total opacity and the $P_{j}$ term in Eq.\,(\ref{eq:rho-main}) is independent from the radiation field, we obtain a system of linear differential equations which can be solved. However, the computational cost of the direct solution is high, and the iterative method is applied in paper \cite{Uitenbroek01} to solve the equation. The following iterative scheme is used:

\vskip-.6cm
\begin{gather}
\vec{n} \nabla I^{(m)}(\nu,\vec{n}) = -\chi ^{tot}I^{(m)}(\nu,\vec{n}) + \frac{h\nu}{4\pi} \sum\limits^{PRD}_{i,j}n_{j}A_{ji}\phi _{ij}\rho _{ij}[I^{(m-1)}(\nu,\vec{n})] + \nonumber \\
\label{eq:PRD-iter-main}
\sum\limits^{other}_{i,j} n_{j}U_{ji} + \eta ^{c}
\end{gather}

We can introduce the discrepancy from the exact solution for the intensity as $\delta I^{(m)} = I - I^{(m)}$, and subtract Eq.\,(\ref{eq:PRD-iter-main}) from Eq.\,(\ref{eq:RT-ndir}). We obtain:

\vskip-.6cm
\begin{gather}
\vec{n} \nabla \delta I^{(m)} = -\chi ^{tot}\delta I^{(m)} + \frac{h\nu}{4\pi} \sum\limits^{PRD}_{i,j}n_{j}A_{ji}\phi _{ij} Q _{ij}[\delta I^{(m-1)}]
\end{gather}

where $Q _{ij}[\delta I^{(m-1)}] = \rho _{ij} [I] - \rho _{ij} [I^{(m-1)}] = \rho _{ij} [I^{(m-1)} + \delta I^{(m-1)}] - \rho _{ij} [I^{(m-1)}]$. Finally, we obtained the iterative scheme to get the radiative field for the populations at each intermediate step of the main iteration loop. From the radiation field we can obtain $\rho _{ij}$, and move to the next step of the main iteration loop. This procedure describes the scheme of Uitenbroek~\cite{Uitenbroek01} for inclusion of the PRD effects into the Rybicki and Hummer \cite{Rybicki92} preconditioning iterative scheme to solve the RT equation.

\end{appendices}

\end{document}